\documentclass[prd,amsfonts,aps,epsf]{revtex4}
\usepackage{graphics}
\usepackage{graphicx}
\usepackage{epsfig}

\begin{document}
\input epsf.tex

\title{A New Template Family For The Detection Of Gravitational Waves From Comparable Mass Black Hole Binaries.}
\author{Edward K. Porter}

\address{Max Planck Institut f\"{u}r Gravitationsphysik, Albert Einstein Institut,\\ Am M\"{u}hlenberg 1, D-14476, Golm bei Potsdam, Germany}

\vspace{1cm}
\begin{abstract}
In order to improve the phasing of the comparable-mass waveform as we approach the last stable orbit for a system, various re-summation methods have been used to improve the standard post-Newtonian waveforms.  In this work we present a new family of templates for the detection of gravitational waves from the inspiral of two comparable-mass black hole binaries.  These new adiabatic templates are based on re-expressing the derivative of the binding energy and the gravitational wave flux functions in terms of shifted Chebyshev polynomials.  The Chebyshev polynomials are a useful tool in numerical methods as they display the fastest convergence of any of the orthogonal polynomials.  In this case they are also particularly useful as they eliminate one of the features that plagues the post-Newtonian expansion.  The Chebyshev binding energy now has information at all post-Newtonian orders, compared to the post-Newtonian templates which only have information at full integer orders.  In this work, we compare both the post-Newtonian and Chebyshev templates against a fiducially exact waveform.  This waveform is constructed from a hybrid method of using the test-mass results combined with the mass dependent parts of the post-Newtonian expansions for the binding energy and flux functions.  Our results show that the Chebyshev templates achieve extremely high fitting factors at all PN orders and provide excellent parameter extraction.  We also show that this new template family has a faster Cauchy convergence, gives a better prediction of the position of the Last Stable Orbit and in general recovers higher Signal-to-Noise ratios than the post-Newtonian templates.\\

\end{abstract}

\maketitle

\section{Introduction}
It is believed that stellar mass compact binaries consisting of double neutron stars (NS-NS), double black holes (BH-BH) or a mixed binary consisting of a neutron star and a black hole (BH-NS), are  the primary targets for a direct first detection of gravitational waves (GW) by the ground-based interferometers, LIGO \cite{LIGO}, VIRGO\cite{VIRGO},  GEO600 \cite{GEO}, and TAMA~\cite{TAMA}.  It is also believed that the inspiral of comparable-mass Supermassive BHs during the merger of galaxies will be a major source of GWs for the proposed space-based detector LISA~\cite{LISA}.    Under radiation reaction the orbit of a binary slowly decays, emitting a signal whose amplitude and frequency increases with time and is termed a ``chirp'' signal. While stellar models predict that there is a greater population of NS-NS binaries \cite{Grish,Phin,Narayan,Stairs,KalogeraAndBelczynski},  it is believed that BH-BH binaries will be the strongest candidates for detection since they can be seen within a greater volume, about two orders-of-magnitude greater than that for NS-NS binaries \cite{PostnovEtAl,Grish}.  A quick calculation shows that the idealized Signal-to-Noise ratio (SNR) at 100 Mpc is $\sim 2.5$ for a NS-NS binary, $\sim 5$ for a BH-NS binary and $\sim 12$ for a BH-BH binary, assuming the LIGO ground-based detector has a low frequency cutoff of 40 Hz.  For LISA, it has been shown that we should, theoretically, be able to view the merger of Supermassive BHs out to cosmological distances of $z\sim 10$~\cite{cornishporter, hughes}.

At present, the best proposed method to detect these sources is the method of matched filtering~\cite{Helst}.  In this method, a set of waveforms or templates, that depend on a number of parameters of the source and its location and orientation relative to the detector, are created. These templates are then cross-correlated with the detector output weighted by the inverse of the noise spectral density.  If a signal, whose parameters are close to one of the template waveforms, is present in the detector output then the cross-correlation builds up.  In the case of a sufficiently strong signal, the correlation will be much larger than the root-mean-squared (RMS) correlation in the absence of any signal.  The success of matched filtering depends on how well we understand the phase evolution of the waveform. A tiny instantaneous difference,  as low as one part in $10^3$, between the phase of the true signal in the detector output and the search template, could lead to a cumulative difference of several radians as we integrate over several hundred to several thousand cycles. With the aim of improving the detection ratio for binary inspirals, there has been a considerable effort in accurately computing the dynamics of a compact binary and the emitted waveform.

There has been a number of parallel efforts using different schemes to accurately describe the phase. Firstly, there is the post-Newtonian (PN) expansion of Einstein's equations which has been used to treat the dynamics of two bodies of comparable masses, with and without spin, in orbit around each other. This approximation is applicable when the velocities involved in the system are small but there is no restriction on the ratio of the masses \cite{BDIWW,BDI,WillWise,BIWW,Blan1,DJSABF,BFIJ}.  Next, black hole perturbation theory has been used to compute the dynamics of a test particle in orbit around a Schwarzschild or Kerr BH. Black hole perturbation theory does not make any assumptions on the velocity of the components, but is valid only in the limit when the mass of one of the bodies is much less than the other \cite{Poisson1,Cutetal1,TagNak,Sasaki,TagSas,TTS}.  Finally, there has been some very promising progress made in the field of numerical relativity~\cite{numrel}.  It is now possible to start numerical simulations a few orbits before merger, and evolve the black holes through the merger and into the ringdown phase.  The goal is to take adiabatic inspiral templates and match them to the numerical waveforms as we approach the highly relativistic phase~\cite{pnnumrel}.  Hopefully, in time, this will provide us with a complete GW template, from the adiabatic regime of the inspiral through the merger and onto the ringdown.

In this work, we are mostly interested in the PN approximation and the adiabatic inspiral phase.  In particular we will focus on the inspiral of BH-BH binaries as these systems will coalesce in the most sensitive frequency band for the ground based detectors.  The PN approximation is a perturbative method which expands the equations of motion, binding energy and GW flux as a power series in $v/c$, where $v$ is the invariant velocity of the system and $c$ is the speed of light.  In the early adiabatic stage of an inspiral, the radiation reaction time-scale is much greater than the orbital time-scale.  At this point in the binary evolution, the velocity of the bodies in the system is quite small.  In fact for a BH-BH system of $(10,10)\,M_{\odot}$, the velocity of the system as it enters the LIGO bandwidth is $\sim 0.23\,c$.  At this point, the two BHs are separated by a distance of $R\sim 19\,m$, where $m$ is the total mass of the system.  This is quite close to the point at $r\leq 10 m$~\cite{Brady} where we know the PN approximation begins to break down.

It was shown~\cite{DIS1, DIS2, portersathya, porter2} that templates based on re-summation methods, such as Pad\'e approximation, have a faster convergence in modelling the gravitational waveform for test-mass systems.  However, in this case an exact expression is known for the binding energy of the system, while the GW flux is known only in terms of a PN expansion in powers of $v$.  When compared with numerical fluxes, the Pad\'e approximations were clearly superior to the standard PN fluxes.  The Pad\'e based templates were then used to partially construct Effective One Body templates~\cite{EOB} which went beyond the adiabatic approximation and modelled the waveform into the merger phase.  Other template families such as the more phenomenological BCV (Buonanno-Chen-Vallisneri) templates~\cite{BCV} and a new family called complete adiabatic~\cite{compad} have also been proposed as possible successors to the PN template.

In this paper, we build on the results of a previous work where we defined a new template family for the Schwarzschild test-mass case that was based on Shifted Chebyshev polynomials (SCPs).  We were able to show that modelling the flux function with these orthogonal polynomials gave a more convergent and robust flux function which leads to lower errors in the estimation of the chirp mass.  In this current work we explore the comparable-mass case.  It is known that the Chebyshev polynomials are bound in the domain between $\pm 1$.  In the previous work, we used the SCPs which we bound between zero and the velocity at the Last Stable Orbit (LSO).  However, it is known in the literature that the smaller the domain we are approximating in, the faster the convergence of a Chebyshev series.  We therefore derive the SCPs which are a function of the total mass of the system.  This ensures that we minimize the domain of the SCPs for different comparable-mass systems to try and ensure the best results.  Using the SCPs we propose a new representation of the binding energy and flux functions.

\subsection{Organization of the Paper.}
In Sec~\ref{sec:waveform} we define the gravitational waveform in the restricted PN approximation for ground based detectors.  We also define PN expansions for the derivative of the binding energy and the flux functions.  In Sec~\ref{sec:Chebyshev} we quickly review some aspects of Approximation theory and outline some of the details of the Chebyshev polynomials.  We then go on to derive the analytic form for the SCPs in terms of the monomials in $v$, as well as analytic expressions for the monomials in terms of the SCPs.  In Sec~\ref{sec:chebwave} we define the new Chebyshev approximations to the binding energy and the GW flux function.  We also define a fiducially exact binding energy and flux which is based on a hybrid scheme that uses aspects from both the test-mass and comparable-mass cases.  Using these fiducial functions, we make a graphical comparison of the performance of both the PN and Chebyshev approximations against the respective fiducial functions.  Sec~\ref{sec:fitfac} contains a more quantitative analysis of the new approximations.  Using the newly defined functions we calculate the fitting factors of the Chebyshev and PN waveforms against a fiducial exact waveform.  In Sec~\ref{sec:cauchy} we investigate the Cauchy convergence of this new template family.  In Secs~\ref{sec:lso} and \ref{sec:snr} we look at how well the various approximations predict the position of the LSO, and how much signal-to-noise ratio they recover as a function of both total mass and distance when compared to the fiducial waveform.

\section{The Gravitational Waveform.}\label{sec:waveform}
In the transverse-traceless gauge GWs are represented by the two polarizations $h_+$ and $h_\times.$ The response $h(t)$ of a detector to an incoming signal is given by the combination $h(t) = h_+ F^{+} + F^{\times} h_\times$, where in the restricted post-Newtonian approximation~\cite{Cutetal2} the polarizations are given by
\begin{eqnarray}
h_+ (t) & = & \frac{4\eta m}{D} \frac{(1+\cos^2 i)}{2}\, v^2(t)\, \cos \phi(t), \\
h_\times(t) & = & \frac{4\eta m}{D}\, \cos i\, v^{2}(t)\, \sin \phi(t),
\label{kt7}
\end{eqnarray}
and the beam-pattern functions are given by
\begin{eqnarray}
F^{+}\left(\theta, \phi, \psi\right) &=& 
\frac{1}{2} \left( 1 + \cos^2 \theta\right) 
\cos 2\phi\cos 2\psi -  \cos \theta \sin 2\phi\sin 2\psi,\\
\label{kt4}
F^{\times}\left(\theta, \phi, \psi\right) &=& 
\frac{1}{2} \left( 1 + \cos^2 \theta\right) 
\cos 2\phi \sin 2\psi +  \cos \theta \sin 2\phi \cos 2\psi.
\label{kt5}
\end{eqnarray}
By restricted PN approximation we retain only the dominant amplitude and ignore all higher harmonic corrections.  In the above equations, $m=m_{1}+m_{2}$ is the total mass of the system, $\eta=m_{1}m_{2}/m^{2}$ is the reduced mass ratio, $i$ is the inclination angle of the binary orbit, $D$ is the distance to the source, $v= (m\Omega)^{1/3}=(\pi m f_{GW})^{1/3}$ is the invariant PN velocity parameter related to the orbital frequency $\Omega$ and GW frequency $f_{GW}$, $(\theta,\phi)$ define the sky position of the source and $\psi$ is the polarization angle of the wave.
As a single detector will be unable to disentangle the two individual polarizations, we can write the response in the form
\begin{equation}
h(t) = \frac{4\eta m}{D} {\cal A}\, v^{2}(t)\, \cos[\phi(t) + \phi_0],
\end{equation}
where, for the short-duration signals we are considering in this study, the coefficient ${\cal A}={\cal A}(\theta,\phi, i,\psi)$ and phase factor $\phi_0=\phi_{0}(\theta,\phi, i,\psi)$ can be taken to be constant.

The PN approximation and the quadrupole formula, when applied to an inspiralling binary system, give the relativistic binding energy per unit mass $E(v)$, and the GW flux $F(v)$ as series expansions in the parameter $v.$  Once we have these two functions we can use the energy balance argument 
\begin{equation}
m\,\frac{dE(v)}{dt} = -F(v)
\end{equation}
to obtain the evolution of the phase of the GW.  Integrating the energy balance equation and using $2\pi f = d\phi/dt,$ we obtain the set of parametrized integral equations
\begin{eqnarray}
t(v) & = & t_{0} + m \int_{v}^{v_{lso}} dv \, \frac{{E}'(v)}{F(v)},
\label{eq:timeVsv} \\
\phi(v) & = & \phi_{0} + 2 \int_{v}^{v_{lso}} dv\, v^3 \, \frac{{E}'(v)}{F(v)},
\label {eq:phiVsv}
\end{eqnarray}
where $E'(v)=dE(v)/dv,$ $t_{0}$ and $ \phi _{0}$ is can be taken to be constants of integration at a particular reference time.  Rather than solving the set of equations by numerical integration, it is much quicker to solve the following set of ordinary differential equations (ODEs)
\begin{equation}
\frac{dv}{dt} = -\frac{F(v)}{mE'(v)}, \ \ \ \ 
\frac{d\phi}{dt} = \frac{2 v^{3}}{m}.
\label{eqn:odes}
\end{equation}
The above integral equations and ODEs for the phasing formula $\phi = \phi(t)$ hold under the `adiabatic inspiral' assumption.

While for test-mass systems we have an exact expression for the binding energy and a PN expansion for the flux function, we have no such luxury in this case.  For comparable mass bodies, the binding energy is represented by a PN series approximation to order ${\mathcal O}(v^{6})$, while the GW flux function is again represented by a series approximation to order ${\mathcal O}(v^{7})$.  Thus we replace the functions $(E'(v), F(v))$ in Eqns~(\ref{eqn:odes}) with the PN power series approximations $(E_{T_{n}}'(v), F_{T_{n}}(v))$.

The PN energy derivative is defined by the power series expansion~\cite{e1, e2, e3, e4, e5}
\begin{equation}
E_{T_{n}}'(v) = -\eta v \sum_{k=0}^{6} e_{k}v^{k} 
\label{eqn:pnenergy}
\end{equation}
where
\begin{eqnarray}
e_{0} & = & 1\,\,\,\,\,\, , \,\,\,\,\,\, e_{2} =  -\frac{1}{6}\left(9+\eta\right)\nonumber\\ \nonumber \\
e_{4} & = & -\frac{3}{8}\left(27-19\eta+\frac{\eta^{2}}{3}\right)\nonumber\\ \nonumber \\
e_{6} & = & 4\left(-\frac{675}{64} +\left[\frac{34445}{576} - \frac{205\pi^{2}}{96}\right]\eta -\frac{155}{96}\eta^{2}-\frac{35}{5184}\eta^{3}\right)
\end{eqnarray}
and $e_{1} = e_{3} = e_{5} = 0$.   In the coefficient $e_{6}$ we have used the fact that the constant term appearing in the literature has been determined to be $\lambda=-1987/3080$~\cite{e3, e4, e6, e6}, which allows us to write the terms in square brackets in a more condensed form.

The GW flux function, also defined in the form of a PN expansion, is given by~\cite{BDIWW,BDI,WillWise,BIWW,Blan1,DJSABF,BFIJ}
\begin{equation}
F_{T_{n}}(v) = F_{N}\left[\sum_{k=0}^{7}A_{k}(\eta)v^{k} + \ln(v)B_{6}v^{6} \right],
\label{eqn:pnflux}
\end{equation}
where the Newtonian flux is given by
\begin{equation}
F_{N}(v) = \frac{32}{5}\eta^{2}v^{10}, 
\end{equation}
and
\begin{eqnarray}
A_{0} &= & 1 \,\,\,\,\,\, , \,\,\,\,\,\,A_{1} = 0 \,\,\,\,\,\, , \,\,\,\,\,\,A_{2}  = -\frac{1247}{336} - \frac{35}{12}\eta\,\,\,\,\,\, , \,\,\,\,\,\,A_{3}  =  4\, \pi,\nonumber\\ \nonumber\\
A_{4} & = & -\frac{44711}{9072} + \frac{9271}{504}\eta + \frac{65}{18}\eta^{2},\nonumber\\ \nonumber \\
A_{5} & = & -\left(\frac{8191}{672} +\frac{583}{24}\eta\right)\pi,\nonumber\\ \nonumber\\
A_{6} & = &\frac{6643739519}{69854400} - \frac{1712\, \gamma}{105} + \frac{16\, \pi^{2}}{3} - \frac{3424\, \ln(2)}{105}-  \left(\frac{41\pi^{2}}{48}-\frac{134543}{7776}\right)\eta -\frac{94403}{3024}\eta^{2} -\frac{775}{324}\eta^{3} ,\nonumber\\ \nonumber\\
A_{7}& = &\pi\left(-\frac{16285}{504} + \frac{214745}{1728}\eta + \frac{193385}{3024}\eta^{2} \right),\\ \nonumber\\
B_{6} &=&-1712/105,
\end{eqnarray}
where $\gamma=0.577...$ is Euler's constant.  Once again, in coefficient $A_{6}$ we have used the value for the constant $\Theta = -11831/9240$~\cite{fconst} that appears in the literature to write the $\eta$ dependent term in a shortened form.  We can see that the flux is not defined by a pure power series, but has a logarithmic term appearing at the 3-PN order.  Terms like this can be responsible for poor convergence if not treated properly.

\section{The Shifted Chebyshev Polynomials and The Velocity Monomials}\label{sec:Chebyshev}
As we have seen, both the binding energy and the GW flux functions are given by power series representations to various orders.  We can treat these expansions just like we would a standard Taylor series.  We should caution the reader not to confuse the subscript $T_{n}$ with the symbol for the Chebyshev polynomial $T_{n}(x)$ that we will define in this section.  While it is an unfortunate coincidence, we will continue to use both symbols as they are established in the literature.  We have seen in other studies~\cite{DIS1, DIS2,  portersathya, porter2} that in the case of a test-mass particle orbiting both Schwarzschild and Kerr black holes, the PN representation of the flux does not display the fastest convergence.  Other re-summation methods such as Pad\'e approximation offer a faster converging approximation to the GW flux function when compared with numerical results from Black Hole Perturbation Theory.  It is these previous works that motivate us to try a different and more convergent approach to the modelling of the binding energy and flux functions.

The problem for expansions like a Taylor or a Maclaurin series is that they are based on the Weierstrass theorem, i.e. {\em For any continuous function $f(x)\in{\mathcal C}[a,b]$ and for any given $\epsilon > 0$, there exists a polynomial $p_{n}(x)$ for some sufficiently large $n$ such that $|f(x)-p_{n}(x)|_{max}<\epsilon$}.  So, as long as we can approximate a function with a series containing a sufficiently large number of terms, we can always minimize the error of our approximation.  However, this is not always practical as the number of necessary terms may be too high (for example, it takes about 5000 terms in a Taylor expansion of $arctan(x)$ to deliver a five significant figure accuracy at $x$ equal to unity~\cite{FSActon}). Moreover, we may be dealing with an approximation to some function where it is very difficult to calculate more than a few terms.  For comparable mass systems this is a particular problem as we have only a three term expression for the binding energy and a seven term expansion for the GW flux function. We know from previous studies in the test-mass case that this number of terms may not be sufficient to accurately model the true binding energy and flux functions.  A more promising possibility is based on the Chebyshev Alternation theorem for polynomials which states `{\em For any continuous function $f(x)\in{\mathcal C}[a,b]$, a unique minimax polynomial approximation $p_{n}(x)$ exists, and is uniquely characterized by the `alternation property' (or `equioscillation property') that there are (at least) $n+2$ points in $[a,b]$ at which $|f(x)-p_{n}(x)|$ attains its maximum absolute value with alternating signs}'.  Thus, the reason the minimax polynomial is so sought after is that the Chebyshev Alternation theorem guarantees that there is only one such polynomial and it has the necessary condition of having an  `equal-ripple' error curve. Unfortunately, the minimax polynomial is usually extremely difficult, if not impossible, to find. A more promising and obtainable possibility is based on getting close to the minimax polynomial by using the family of Ultraspherical (or Gegenbauer) polynomials which are defined by
\begin{equation}
P_{n}^{(\alpha)}(x)=C_{n}\left(1-x^{2}\right)^{-\alpha}\frac{d^{n}}{dx^{n}}\left(1-x^{2}\right)^{n+\alpha}\,\,\,\,\,\,\,\,\,\,\,\,\,\left(-1\leq\alpha\leq\infty\right),
\end{equation}
where $C_{n}$ is a constant.  These polynomials are orthogonal over $x\in[-1,1]$ with respect to the weight function $\left(1-x^{2}\right)^{\alpha}$.  A feature of the polynomials $P_{n}^{(\alpha)}(x)$ is that they have $n$ distinct and real zeros and exhibit an oscillatory behavior in the interval $[-1,1]$.  For the particular value of $\alpha=-1/2$ the amplitude of the oscillations remain constant throughout the interval and is conducive to finding an "equal-ripple" error curve, which is integral to the minimax polynomial.

This value of $\alpha$ corresponds to the Chebyshev polynomials of the first kind $T_{n}(x)$ (hereafter Chebyshev polynomials).  These polynomials are closely related to the minimax polynomial due to the fact that there are $n+1$ points in [-1,1] where $T_{n}(x)$ attains a maximum value of unity with alternating signs~\cite{Mason}.  It can be shown~\cite{Snyder} that the Chebyshev polynomials exhibit the fastest convergence properties of all of the Ultraspherical polynomials. The $n+1$ extrema are given by
\begin{equation}
x_{k}=\cos\left(\frac{k\pi}{n}\right)\,\,\,\,\,\,\,\,\,\,\,\,\,\,\,\,\,\,\,k=0,..,n ,
\end{equation}
with $n$ zeros given by
\begin{equation}
x_{k}=\cos\left(\frac{[k-\frac{1}{2}]\pi}{n}\right)\,\,\,\,\,\,\,\,\,\,\,\,\,\,\,\,\,\,\,k=1,..,n .
\end{equation}
The Chebyshev polynomials are formally defined by 
\begin{equation}
T_{n}(x) = \cos(n\theta)\,\,\,\,\,\,\,\,\,\,\,\,\,\,\,\,\,\,\,when\,\,\,x=\cos(\theta).
\end{equation}
From de Moivre's theorem, we know that $\cos\left(n\theta\right)$ is a polynomial of degree $n$ in $\cos\left(\theta\right)$, e.g. 
\begin{equation}
\cos\left(0\theta\right)= 1,\,\,\,\,\,\,\,\, \cos\left(\theta\right) = \cos\left(\theta\right),\,\,\,\,\,\,\,\, \cos\left(2\theta\right) = 2\cos^{2}\left(\theta\right) -1, .....\;\;,
\end{equation}
which allows us to write the Chebyshev polynomials
\begin{equation}
T_{0}(x) =1,\,\,\,\,\,\,\,\,  T_{1}(x) =x,\,\,\,\,\,\,\,\, T_{2}(x) =2x^{2}-1,.....\;\;.
\end{equation}
Therefore, a Chebyshev series in $x$ corresponds to a Fourier series in $\theta$.  This close relation to a trigonometric function means these polynomials are extremely useful in approximating functions.  The Chebyshev polynomials are orthogonal polynomials with respect to the weight $\left(1-x^{2}\right)^{-1/2}$ according to 
\begin{equation}
\left<T_{i}(x)\left|T_{j}(x)\right.\right> = \int_{-1}^{1}\,dx\frac{T_{i}(x)T_{j}(x)}{\sqrt{1-x^{2}}}=\left\{ \begin{array}{ll} 0 & i\neq j \\ \frac{\pi}{2} & i=j\neq0 \\ \pi & i=j=0 \end{array}\right.
\end{equation}
with initial conditions
\begin{equation}
T_{0}(x)=1\,\,\,\,\,,\,\,\,\,\,T_{1}(x) = x.
\end{equation}
We calculate the higher order Chebyshev polynomials using the recurrence relation
\begin{equation}\label{eqn:reqeqn1}
T_{n}(x)=2xT_{n-1}(x)-T_{n-2}(x),
\end{equation}
given the above initial conditions. 

\begin{figure}[t]
\vspace{0.25 in}
\begin{center}
\centerline{\epsfxsize=10cm \epsfysize=7cm \epsfbox{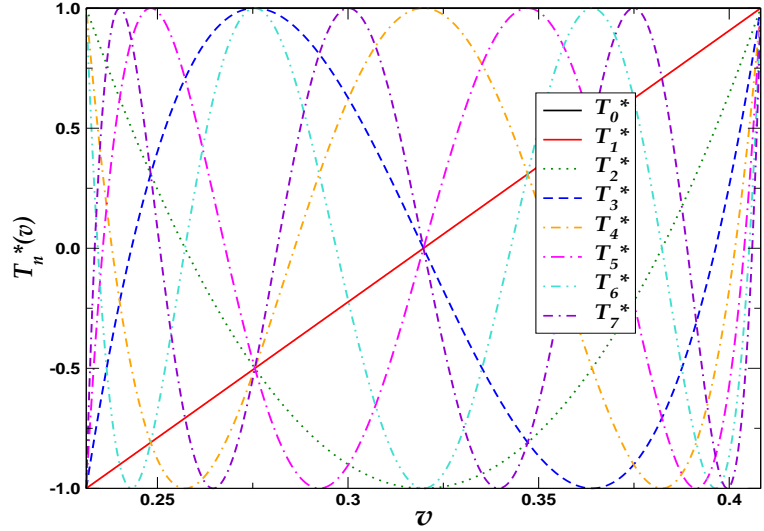}}
\vspace{5mm}
\caption{The 7th order Shifted Chebyshev Polynomials for a $(10,10)\,M_{\odot}$ BH-BH binary in the domain $v\in[0.23,0.408]$.  We can see that the polynomials have $n+1$ alternating maximum and minimum values of $\pm 1$ and $n$ zeros across the domain.}
\label{fig:scp}
\end{center}
\end{figure}

The Chebyshev polynomials are usually defined in the domain [-1,1].  For this particular problem, we require the Shifted Chebyshev Polynomials, $T_{n}^{*}(v)$, which are defined in the domain $v\in [v_{ini},v_{lso}]$ so that we have the maximum convergence possible.  We can transform from the domain [-1,1] to another interval $[a,b]$ using
\begin{equation}
s=\frac{2x-(a+b)}{b-a}\,\,\,\,\,\,\,\,\,\,\,\,\,\,\,\ \forall\,\, x\in[a,b], s\in[-1,1].
\end{equation}
Now first defining the two variables
\begin{equation}
 \chi = v_{ini} + v_{lso}\,\,\,\,\,\,\, , \,\,\,\,\,\,\,\xi = v_{lso} - v_{ini},
\end{equation}
where $v_{ini}=(\pi m f_{low})^{1/3}$ is the initial velocity of the body as it crosses the lower frequency cutoff of the detector and $v_{lso}$ is the velocity at the last stable circular orbit,  the shift has the form 
\begin{equation}
s =  \frac{2v - \chi}{\xi}  \;\;\;\;\;\;\forall\,\, v\in[v_{ini},v_{lso}].
\end{equation}
We should point out here that both $\chi$ and $\xi$ are functions of the total mass of the binary through the initial velocity.  This in turn infers that the Shifted Chebyshev polynomials (SCPs) are also functions of the total mass.  Furthermore, while the PN approximations to the energy and flux functions are unchanged for all equal mass systems, regardless of total mass, this will not be the case for the Chebyshev approximations.  For each equal mass system with a different total mass, the approximations to the energy and the flux will differ.  We now write the SCPs in the form
\begin{equation}
T_{n}^{*}(v)=T_{n}(s)=T_{n}\left(\frac{2v - \chi}{\xi}\right),
\end{equation}
and the recurrence relation as 
\begin{equation}\label{eqn:rec}
T_{n}^{*}(v)=2\left(\frac{2v - \chi}{\xi}\right)T_{n-1}^{*}(v)-T_{n-2}^{*}(v).
\end{equation}
While for computational purposes we will use the above recurrence relation to calculate the shifted polynomials, we also need to be able to express the polynomials in analytical form.  Once we have the analytic expressions for $T_{n}^{*}(v)$ in terms of $v$, we can then invert the expressions to find an analytic format for the monomials of $v$ in terms of $T_{n}^{*}(v)$.  Therefore, using the fact that the SCPs are normalized according to 
\begin{eqnarray}
T_{0}^{*}(v) &=& 1\\ \nonumber \\
T_{1}^{*}(v) &=& \xi^{-1}\left[2v-\chi\right]
\end{eqnarray}
we can use the recurrence relation to find the other shifted polynomials to order ${\mathcal O}(v^{7})$, i.e.
\begin{eqnarray}
T_{2}^{*}(v) &=& \xi^{-2}\left[\left(2\chi^{2}-\xi^{2}\right) - 8\chi v - 8v^{2}\right]\\ \nonumber \\
T_{3}^{*}(v) &=& \xi^{-3}\left[\left(3\chi\xi^{2}-4\chi^{3}\right)+6\left(4\chi^{2}-\xi^{2}\right)v-48\chi v^{2}+32v^{3}\right]\\  \nonumber \\
T_{4}^{*}(v) &=& \xi^{-4}\left[\left(\xi^{4}+8\chi^{4}-8\chi^{2}\xi^{2}\right)+32\left(\chi^{2}\xi^{2}-2\chi^{3}\right)v + 32\left(6\chi^{2}-\xi^{2}\right)v^{2}-256\chi v^{3}+218v^{4}\right] \\  \nonumber \\
T_{5}^{*}(v) &=&\xi^{-5}\left[\left(20\chi^{3}\xi^{2}-16\chi^{5}-5\chi\xi^{4}\right) + \left(160\chi^{4}-120\chi^{2}\xi^{2}+10\xi^{4}\right)v + \left(240\chi\xi^{2}-640\chi^{3}\right)v^{2}\right.\nonumber \\ 
&+&\left.\left(240\chi\xi^{2}-640\chi^{3}\right)v^{3} -1280\chi v^{4}+ 512v^{5}\right]\\  \nonumber \\
T_{6}^{*}(v) &=& \xi^{-6}\left[\left(32\chi^{6} - 48\chi^{4}\xi^{2} + 18\chi^{2}\xi^{4} - \xi^{6}\right)+\left(384\chi^{3}\xi^{2}-384\chi^{5}-72\chi\xi^{4}\right)v + \left(1920\chi^{4} - 1152(\chi\xi)^{2}+72\xi^{4}\right)v^{2}\right.\nonumber\\
&+&\left.\left(1536\chi\xi^{2}-5120\chi^{3}\right)v^{3} + \left(7680\chi^{2}-768\xi^{2}\right)v^{4}-6144\chi v^{5} + 2048v^{6}\right]\\  \nonumber \\
T_{7}^{*}(v) &=& \xi^{-7}\left[\left(112\chi^{5}\xi^{2}-64\chi^{7}-56\chi^{3}\xi^{4}+7\chi\xi^{6}\right)+\left(896\chi^{6}-1120\chi^{4}\xi^{2}+336\chi^{2}\xi^{4}-14\xi^{6}\right)v\right.\nonumber \\ 
&+&\left.\left(4480\chi^{3}\xi^{2}-5376\chi^{5}-672\chi\xi^{4}\right)v^{2}+\left(17920\chi^{4}-8960(\chi\xi)^{2}+448\xi^{4}\right)v^{3}+\left(8960\chi\xi^{2}-35840\chi^{3}\right)v^{4}\right.\nonumber \\
&+&\left.\left(43008\chi^{2}-3584\xi^{2}\right)v^{5}-28672\chi v^{6} + 8192v^{7}\right]
\end{eqnarray}
Using the above expressions we can now write the monomials of $v$ in terms of $T_{n}^{*}(v)$ as follows
\begin{eqnarray}
v &=& \frac{1}{2}\left(\chi T_{0}^{*}+\xi T_{1}^{*}\right)\\ \nonumber \\
v^{2} &=& \left(\frac{\xi^{2}}{8}+\frac{\chi^{2}}{4}\right)T_{0}^{*} +\frac{\chi\xi}{2}T_{1}^{*} + \frac{\xi^{2}}{8}T_{2}^{*}\\ \nonumber \\
v^{3} &=& \left(\frac{3}{16}\chi\xi^{2}+\frac{\chi^{3}}{8}\right)T_{0}^{*} + \left(\frac{3}{8}\chi^{2}\xi+\frac{3}{32}\xi^{3}\right)T_{1}^{*} + \frac{3}{16}\chi\xi^{2}T_{2}^{*} + \frac{\xi^{3}}{32}T_{3}^{*}\\ \nonumber\\
v^{4} &=&  \left(\frac{3}{128}\xi^{4}+\frac{\chi^{4}}{16}+\frac{3}{16}\chi^{2}\xi^{2}\right)T_{0}^{*} + \left(\frac{1}{4}\chi^{3}\xi+\frac{3}{16}\chi\xi^{3}\right)T_{1}^{*} + \left(\frac{3}{16}\chi^{2}\xi^{2}+\frac{\xi^{4}}{32}\right)T_{2}^{*} + \frac{1}{16}\chi\xi^{3}T_{3}^{*} + \frac{\xi^{4}}{128}T_{4}^{*}\\ \nonumber \\
v^{5} &=& \left(\frac{15}{256}\chi\xi^{4} + \frac{5}{32}\chi^{3}\xi^{2} + \frac{\chi^{5}}{32}\right)T_{0}^{*} +\left(\frac{5}{256}\xi^{5}+\frac{15}{64}\chi^{2}\xi^{3}+\frac{5}{32}\chi^{4}\xi\right)T_{1}^{*} +\left(\frac{5}{32}\chi^{3}\xi^{2}+\frac{5}{64}\chi\xi^{4}\right)T_{2}^{*} \nonumber\\
&+& \left(\frac{5}{64}\chi^{2}\xi^{3} + \frac{5}{512}\xi^{5}\right)T_{3}^{*} + \frac{5}{256}\chi\xi^{4}T_{4}^{*} + \frac{\xi^{5}}{512}T_{5}^{*} \\ \nonumber \\
v^{6} &=& \left(\frac{\chi^{6}}{64}+\frac{15}{128}\chi^{4}\xi^{2}+\frac{45}{512}\chi^{2}\xi^{4}+\frac{5}{1024}\xi^{6}\right)T_{0}^{*} + \left(\frac{15}{256}\chi\xi^{5}+\frac{15}{64}\chi^{3}\xi^{3}+\frac{3}{32}\chi^{5}\xi\right)T_{1}^{*} \nonumber\\
&+& \left(\frac{15}{2048}\xi^{6}+\frac{15}{128}\chi^{2}\xi^{4}+\frac{15}{128}\chi^{4}\xi^{2}\right)T_{2}^{*} + \left(\frac{5}{64}\chi^{3}\xi^{3}+\frac{15}{512}\chi\xi^{5}\right)T_{3}^{*} +\left(\frac{3}{1024}\xi^{6}+\frac{15}{512}\chi^{2}\xi^{4}\right)T_{4}^{*} \nonumber \\
&+& \frac{3}{512}\chi\xi^{5}T_{5}^{*} + \frac{\xi^{6}}{2048}T_{6}^{*} \\ \nonumber \\
v^{7} &=& \left(\frac{\chi^{7}}{128}+\frac{35}{2048}\chi\xi^{6}+\frac{21}{256}\chi^{5}\xi^{2}+\frac{105}{1024}\chi^{3}\xi^{4}\right)T_{0}^{*} +\left(\frac{105}{1024}\chi^{2}\xi^{5}+\frac{7}{128}\chi^{6}\xi+\frac{105}{512}\chi^{4}\xi^{3}+\frac{35}{8192}\xi^{7}\right)T_{1}^{*} \nonumber\\
&+& \left(\frac{21}{256}\chi^{5}\xi^{2}+\frac{35}{256}\chi^{3}\xi^{4}+\frac{105}{496}\chi\xi^{6}\right)T_{2}^{*} +\left(\frac{105}{2048}\chi^{2}\xi^{5}+\frac{35}{512}\chi^{4}\xi^{3}+\frac{21}{8192}\xi^{7}\right)T_{3}^{*} \nonumber\\
&+& \left(\frac{21}{2048}\chi\xi^{6}+\frac{35}{1024}\chi^{3}\xi^{4}\right)T_{4}^{*} + \left(\frac{21}{2048}\chi^{2}\xi^{5}+\frac{7}{8192}\xi^{7}\right)T_{5}^{*} +\frac{7}{4096}\chi\xi^{6}T_{6}^{*} + \frac{\xi^{7}}{8192}T_{7}^{*} 
\end{eqnarray}
To express the PN expansions for both the energy derivative and the gravitational flux function in terms of shifted polynomials, we will substitute the above expressions for the monomials in $v$ and collect all terms proportional to each polynomial.  In Fig~(\ref{fig:scp}) we plot the SCPs for a $(10,10)\,M_{\odot}$ binary BH.  We can see that the polynomials are bound between [-1,1] for  $v\in[0.23,0.408]$ which correspond to the initial and LSO velocities for this particular system assuming a lower frequency cutoff of 40 Hz and that the velocity at the LSO is given by the test-mass Schwarzschild value of $v_{lso}=1/\sqrt{6}$.  We can also see that SCPs have an equal amplitude oscillation exhibiting $n+1$ alternating maximum and minimum values of $\pm 1$, as well as $n$ zeros across the domain.

\section{The Chebyshev Approximation Gravitational Waveform}\label{sec:chebwave}
Our aim in this section is to take the set of ODEs given by Eqn~(\ref{eqn:odes}) and write them in the form
\begin{equation}
\frac{dv}{dt} = -\frac{F_{C_{n}}(v)}{mE_{C_{n}}'(v)}, \ \ \ \ 
\frac{d\phi}{dt} = \frac{2 v^{3}}{m} \label{eq:ODEs}.
\label{eqn:odesc}
\end{equation}
where $E_{C_{n}}'(v)$ and $F_{C_{n}}(v)$ are more robust and convergent expansions of the binding energy and flux functions based on SCPs.

\begin{figure}[t]
\begin{center}
\centerline{\epsfxsize=10cm \epsfysize=6cm \epsfbox{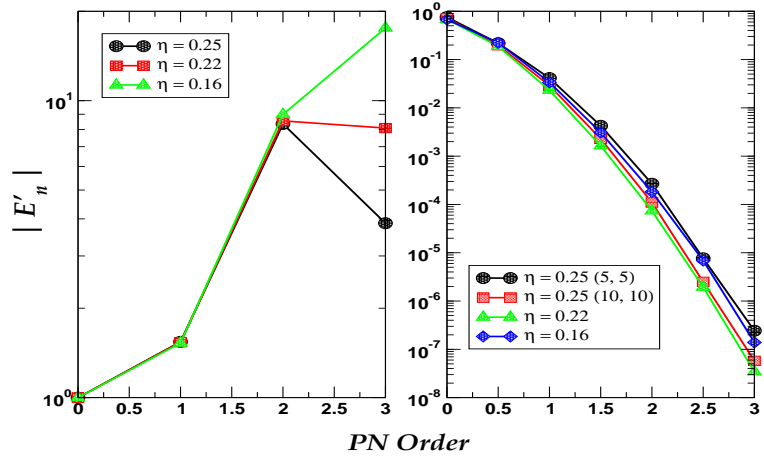}}
\vspace{5mm}
\caption{A comparison of the convergence for the binding energy derivative coefficients of the PN approximation (left) and the shifted Chebyshev approximation (right) for the four BH-BH systems with masses $(5,5),(10,10), (20, 10)$ and $(20,5)\,M_{\odot}$.  We can see that while the values of the PN coefficients oscillate wildly, the Chebyshev coefficients get smaller as we go to higher orders of approximation.  We have plotted two systems with $\eta = 0.25$ to emphasize the fact that the Chebyshev coefficients ${\mathcal E}_{k}={\mathcal E}_{k}(m)$ are a function of the total mass of the system.}
\label{fig:edcoeffcompar}
\end{center}
\end{figure}

\subsection{Modelling The Energy Function.}
We begin re-writing the monomials for $v^{2}, v^{4}$ and $v^{6}$ appearing in Eqn~(\ref{eqn:pnenergy}) in terms of the SCPs. This allows us to  define the Chebyshev approximation to the energy derivative function
\begin{equation}
E_{C_{n}}'(v) = -\eta v \sum_{k=0}^{6} {\mathcal E}_{k}T^{*}_{k}(v), 
\end{equation}
where the new coefficients are defined by
\begin{eqnarray}
{\mathcal E}_{0} & = & 1 + e_{2}\left(\frac{\xi^{2}}{8}+\frac{\chi^{2}}{4}\right) + e_{4}\left(\frac{3}{128}\xi^{4}+\frac{\chi^{4}}{16}+\frac{3}{16}\chi^{2}\xi^{2}\right) + e_{6}\left(\frac{\chi^{6}}{64}+\frac{15}{128}\chi^{4}\xi^{2}+\frac{45}{512}\chi^{2}\xi^{4}+\frac{5}{1024}\xi^{6}\right)\\ \nonumber \\
{\mathcal E}_{1} & = & e_{2}\frac{\chi\xi}{2} + e_{4}\left(\frac{1}{4}\chi^{3}\xi+\frac{3}{16}\chi\xi^{3}\right) + e_{6}\left(\frac{15}{256}\chi\xi^{5}+\frac{15}{64}\chi^{3}\xi^{3}+\frac{3}{32}\chi^{5}\xi\right)\\ \nonumber \\
{\mathcal E}_{2} & = & e_{2}\frac{\xi^{2}}{8} + e_{4}\left(\frac{3}{16}\chi^{2}\xi^{2}+\frac{\xi^{4}}{32}\right) + e_{6}\left(\frac{15}{2048}\xi^{6}+\frac{15}{128}\chi^{2}\xi^{4}+\frac{15}{128}\chi^{4}\xi^{2}\right)\\ \nonumber \\
{\mathcal E}_{3} & = & e_{4}\frac{1}{16}\chi\xi^{3} + e_{6}\left(\frac{5}{64}\chi^{3}\xi^{3}+\frac{15}{512}\chi\xi^{5}\right)\\ \nonumber \\
{\mathcal E}_{4} & = & e_{4}\frac{\xi^{4}}{128} + e_{6}\left(\frac{3}{1024}\xi^{6}+\frac{15}{512}\chi^{2}\xi^{4}\right)\\ \nonumber \\
{\mathcal E}_{5} & = & e_{6}\frac{3}{512}\chi\xi^{5}\\ \nonumber \\
{\mathcal E}_{6} & = & e_{6}\frac{\xi^{6}}{2048}
\end{eqnarray}
Notice that the Chebyshev energy derivative is defined at all PN orders.  The PN approximation suffered from the fact that for any odd PN order template, we had to use the previous order PN approximation for the energy derivative.  This new expression allows us to use an equivalent order approximation to the flux function when we define waveforms at odd PN orders.  We also remind the reader that the function $E_{C_{n}}'(v)=E_{C_{n}}'(v;\eta,m)$ through the Shifted Chebyshev polynomials, whereas $E_{T_{n}}'(v) =E_{T_{n}}'(v;\eta) $ only.

\begin{figure}[t]
\begin{center}
\centerline{\epsfxsize=12cm \epsfysize=9cm \epsfbox{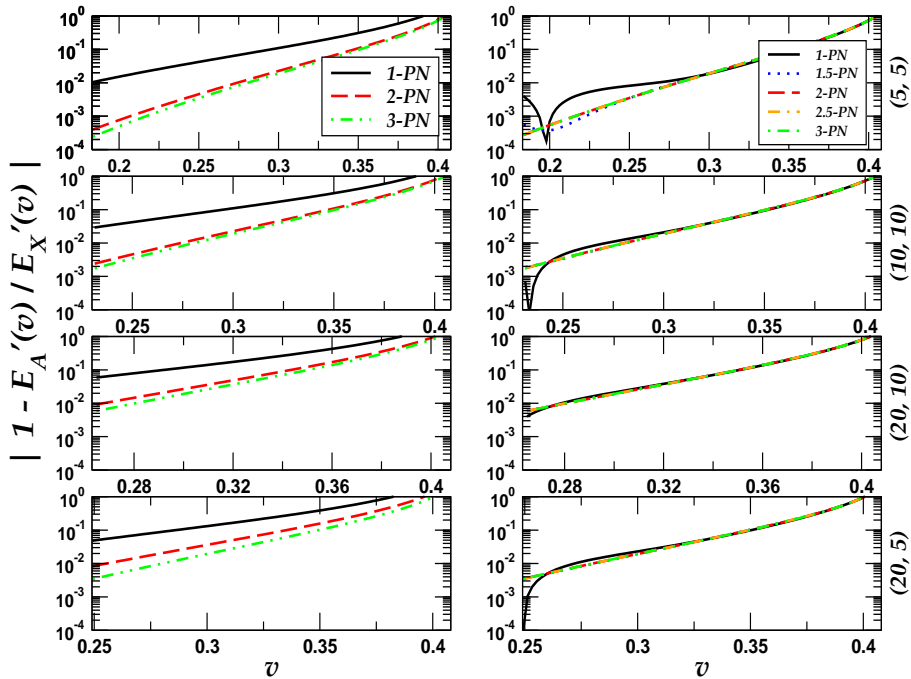}}
\vspace{7mm}
\caption{Comparison of the relative errors for the PN (left) and Chebyshev (right) Energy Derivative functions against a fiducial exact energy function for four BH-BH systems with masses $(5,5),(10,10), (20, 10)$ and $(20,5)\,M_{\odot}$.  Each dip in the approximation error curves correspond to a crossing of the exact energy function by the approximated flux.  The initial velocity in each case is defined by a lower frequency cutoff of 40 Hz.}
\label{fig:edcomparbhbh}
\end{center}
\end{figure}

In Fig~(\ref{fig:edcoeffcompar}) we plot the absolute value of the coefficients from both the PN and Chebyshev approximations for four BH-BH systems with masses $(5,5),(10,10)$ and $(20,10)\,M_{\odot}$ to give values of $\eta$ between 0.16 and 0.25.  We can see from the left hand cell that the PN coefficients display no obvious convergence, with the values of the coefficients growing as we go to higher orders of approximation.  On the other hand, the Chebyshev coefficients display a remarkable convergence with each successive coefficient smaller than the previous as we move to higher PN orders.  This informs us that the Chebyshev series is convergent.  We have shown in a previous work~\cite{portercheb1} that as the shifted polynomials have maximum absolute values of $|T_{n}^{*}(v)|\leq 1$, the value of the coefficients ${\mathcal E}_{n}$ give an excellent estimate of the truncation error at each PN values for the Chebyshev series.

In order to make some kind of comparison we have to define some fiducially exact energy derivative function.  For this particular exercise we define a hybrid~\cite{Willhyb} energy function using the information we have from both the test-mass and comparable-mass cases.  Using the $\eta$ dependent terms from the PN energy we define
\begin{equation}
E_{X}'(v) = -\eta v\left[\frac{1-6v^{2}}{\left(1-3v^{2}\right)^{3/2}}-\frac{\eta}{6}v^{2}-\frac{3}{8}\left(-19\eta+\frac{\eta^{2}}{3}\right)v^{4} 
+\left(4\left[\frac{34445}{576}-\frac{205\pi^{2}}{96}\right]\eta-\frac{155}{24}\eta^{2}-\frac{35}{1296}\eta^{3}\right)v^{6}\right].
\label{eqn:ex}
\end{equation}
where the first term in the brackets is the exact expression for a test-particle in a Schwarzschild geometry.  While we make the assumption that the comparable-mass case is a smooth $\eta$ deformation of the test-mass case, this seems to be a plausible function to use as it returns the test-mass result in the case of $\eta\rightarrow 0$.

In Fig~(\ref{fig:edcomparbhbh}) we plot the relative error between the PN and Chebyshev approximations and the fiducial energy function for a number of different BH-BH systems, i.e.
\begin{equation}
err=\left|1-\frac{E_{A}'(v)}{E_{X}'(v)}\right|.
\end{equation}
We can see from the left hand column that in all cases the PN energy at 1-PN order has an expected high error.  There is then a large jump to the 2-PN order and it is clear that the 3-PN energy gives a slightly better fit than the 2-PN energy.  From the images one could imagine that the PN approximation is converging to a particular solution.  However, one only has to look at the PN approximation for the flux in both the test-mass and comparable-mass cases to realize that if it were possible to go to 4-PN order, it may be that the 4-PN approximation does worse than either the 2 or 3-PN approximations.  On the other hand, we observe a very remarkable convergence with the Chebyshev approximation.  In most cases, the 1-PN order approximation is very close to the performance of the 3-PN Chebyshev approximation.  For the low equal mass case, the approximation has quite a large oscillation at low velocity.  However, this is increasingly minimized as we move to higher mass systems, both equal and unequal-equal.  We can see that in most of the cases on the right hand side, it is virtually impossible to make out the various Chebyshev approximations from 1.5-PN order upwards.

\begin{figure}[t]
\begin{center}
\centerline{\epsfxsize=12cm \epsfysize=7cm \epsfbox{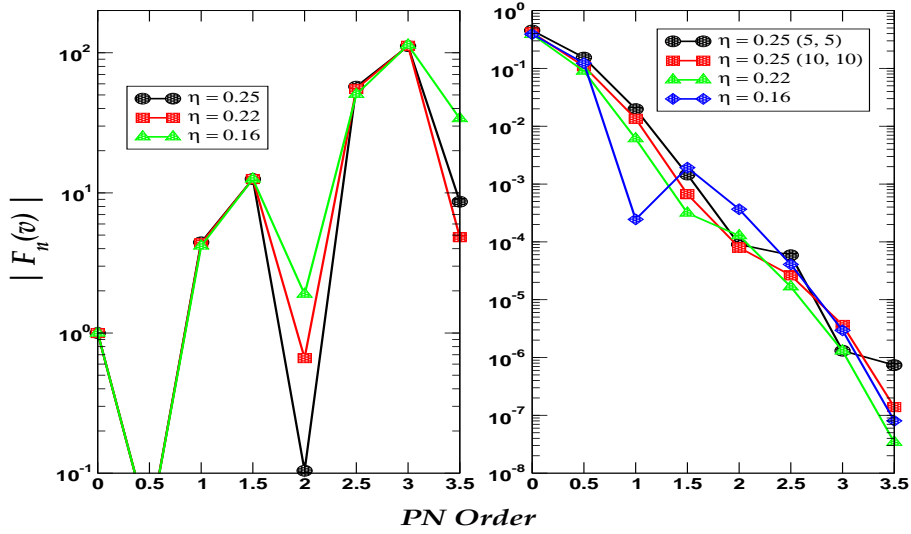}}
\vspace{5mm}
\caption{A comparison of the convergence for the flux coefficients of the PN approximation (left) and the shifted Chebyshev approximation (right) for the four BH-BH systems with masses $(5,5),(10,10), (20, 10)$ and $(20,5)\,M_{\odot}$.  We can see that the values of the PN coefficients again oscillate wildly and display an overall trend of growth.  While not as smooth as in the case of the binding energy, the Chebyshev coefficients again get smaller as we go to higher orders of approximation.}
\label{fig:fluxcoeffcompar}
\end{center}
\end{figure}

\begin{figure}[t]
\begin{center}
\centerline{\epsfxsize=12cm \epsfysize=8cm \epsfbox{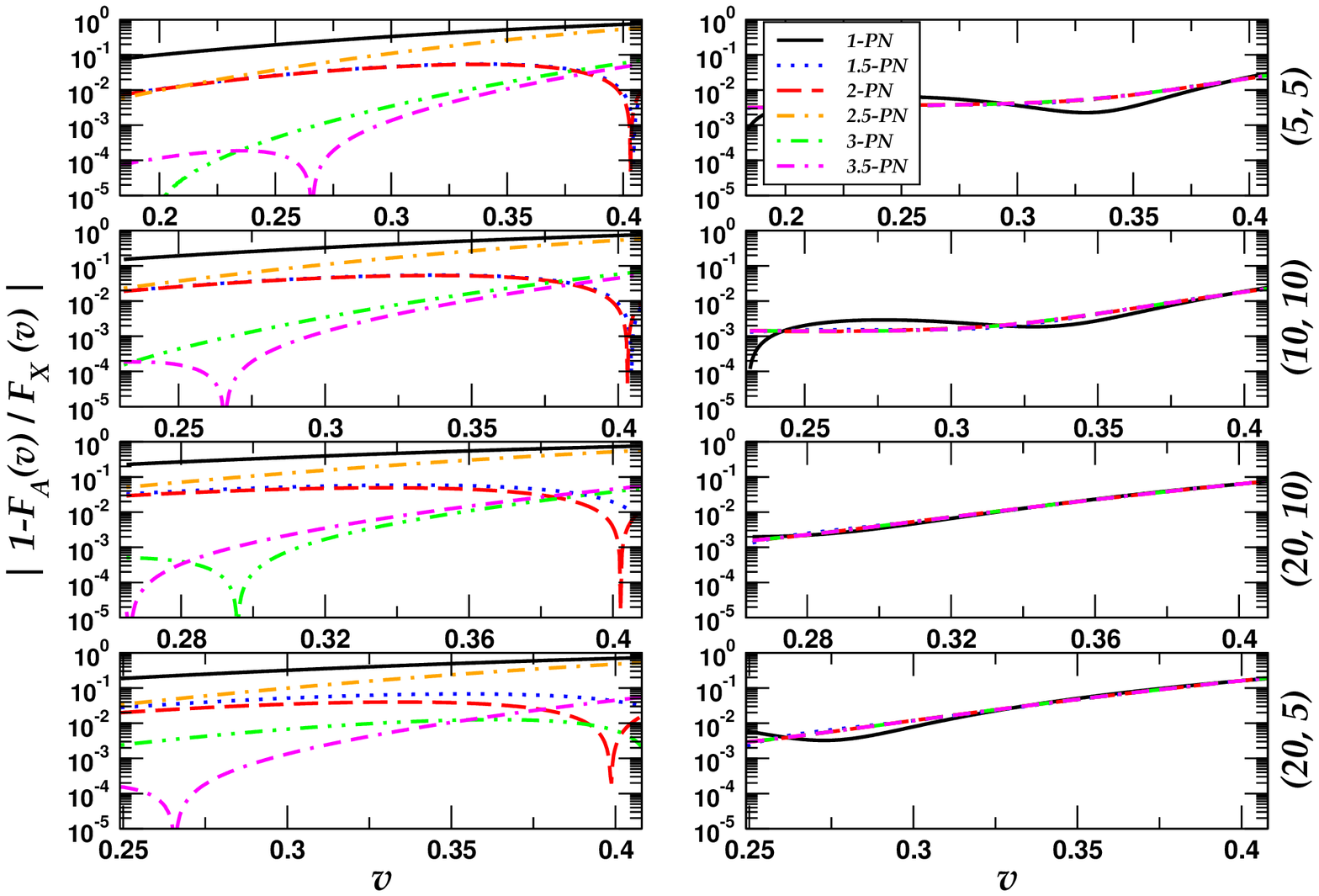}}
\vspace{7mm}
\caption{Comparison of the relative error for the PN (left) and Chebyshev (right) GW flux functions against a fiducial exact flux function for four BH-BH systems with masses $(5,5),(10,10), (20,10)$ and $(20,5)\,M_{\odot}$.  Each dip in the approximation error curve corresponds to a crossing of the exact energy function by the approximated flux.  The initial velocity in each case is defined by a lower frequency cutoff of 40 Hz.}
\label{fig:fluxcomparbhbh}
\end{center}
\end{figure}

\subsection{Modelling The Gravitational Wave Flux Function.}
In order to model the flux function as best as possible, we will borrow some steps from the work on Pad\'e approximation~\cite{DIS1}.  We can see from Eqn~(\ref{eqn:pnflux}) that there is a null linear term in the PN expansion for the flux.  We have confirmed that re-writing the expansion in terms of the SCPs automatically introduces a linear term in the expansion and that compared to the fiducial flux that we will define below, improves the approximation to the flux.  However, we can do better.  In the spirit of the Pad\'e expansion, we first of all define the factored flux function
\begin{equation}
f_{T_{n}} \equiv \left(1-\frac{v}{v_{pole}}\right)F_{T_{n}},
\end{equation}
where $v_{pole} = 1/\sqrt{3}$ is the velocity at the light ring for a test-particle orbiting a Schwarzschild BH.  In order to minimize the effect of the logarithmic term term appearing at 3-PN order, we factor out this term and normalize it according to 
\begin{equation}
f_{T_{n}}(v) = \left[1+\ln\left(\frac{v}{v_{lso}}\right)B_{6}v^{6}\right]
\left[\sum_{k=0}^{7}\,f_{k}v^{k}\right],
\label{eq:facflux}
\end{equation}
where $f_{0} = c_{0}$ and $f_{k} = A_{k} - A_{k-1}/v_{pole},$ for $k = 1,\ldots,7$.  Now once again, swapping all of the monomials in $v$ in the series expansion on the right hand side with their expansions in terms of the SCPs, we define the new Chebyshev flux function
\begin{equation}
F_{C_{n}}(v) = F_{N}\left[1+\ln\left(\frac{v}{v_{lso}}\right)B_{6}v^{6}\right]\left(1-\frac{v}{v_{pole}}\right)^{-1}\sum_{k=0}^{7} {\mathcal F}_{k}T^{*}_{k}(v), 
\end{equation}
where the new coefficients are defined by
\begin{eqnarray}
{\mathcal F}_{0}&=&1+f_{1}\frac{\chi}{2} +f_{2}\left(\frac{\xi^{2}}{8}+\frac{\chi^{2}}{4}\right) +f_{3}\left(\frac{3}{16}\chi\xi^{2}+\frac{\chi^{3}}{8}\right) +f_{4}\left(\frac{3}{128}\xi^{4}+\frac{\chi^{4}}{16}+\frac{3}{16}\chi^{2}\xi^{2}\right) +f_{5}\left(\frac{15}{256}\chi\xi^{4} + \frac{5}{32}\chi^{3}\xi^{2} + \frac{\chi^{5}}{32}\right)\nonumber\\ &+&f_{6}\left(\frac{\chi^{6}}{64}+\frac{15}{128}\chi^{4}\xi^{2}+\frac{45}{512}\chi^{2}\xi^{4}+\frac{5}{1024}\xi^{6}\right) +f_{7}\left(\frac{\chi^{7}}{128}+\frac{35}{2048}\chi\xi^{6}+\frac{21}{256}\chi^{5}\xi^{2}+\frac{105}{1024}\chi^{3}\xi^{4}\right) \\ \nonumber  \\
{\mathcal F}_{1}&=& f_{1}\frac{\xi}{2} + f_{2} \frac{\chi\xi}{2}+ f_{3}\left(\frac{3}{8}\chi^{2}\xi+\frac{3}{32}\xi^{3}\right) + f_{4} \left(\frac{1}{4}\chi^{3}\xi+\frac{3}{16}\chi\xi^{3}\right) + f_{5}\left(\frac{5}{256}\xi^{5}+\frac{15}{64}\chi^{2}\xi^{3}+\frac{5}{32}\chi^{4}\xi\right)\nonumber\\ &+& f_{6}\left(\frac{15}{256}\chi\xi^{5}+\frac{15}{64}\chi^{3}\xi^{3}+\frac{3}{32}\chi^{5}\xi\right) + f_{7}\left(\frac{105}{1024}\chi^{2}\xi^{5}+\frac{7}{128}\chi^{6}\xi+\frac{105}{512}\chi^{4}\xi^{3}+\frac{35}{8192}\xi^{7}\right)\\ \nonumber \\
{\mathcal F}_{2}&=& f_{2}\frac{\xi^{2}}{8} + f_{3}\frac{3}{16}\chi\xi^{2} + f_{4}\left(\frac{3}{16}\chi^{2}\xi^{2}+\frac{\xi^{4}}{32}\right) + f_{5}\left(\frac{5}{32}\chi^{3}\xi^{2}+\frac{5}{64}\chi\xi^{4}\right) + f_{6}\left(\frac{15}{2048}\xi^{6}+\frac{15}{128}\chi^{2}\xi^{4}+\frac{15}{128}\chi^{4}\xi^{2}\right)\nonumber\\ &+& f_{7}\left(\frac{21}{256}\chi^{5}\xi^{2}+\frac{35}{256}\chi^{3}\xi^{4}+\frac{105}{496}\chi\xi^{6}\right)\\ \nonumber \\
{\mathcal F}_{3}&=& f_{3}\frac{\xi^{3}}{32} + f_{4}\frac{\chi\xi^{3}}{16} + f_{5}\left(\frac{5}{64}\chi^{2}\xi^{3} + \frac{5}{512}\xi^{5}\right) + f_{6}\left(\frac{5}{64}\chi^{3}\xi^{3}+\frac{15}{512}\chi\xi^{5}\right) + f_{7}\left(\frac{105}{2048}\chi^{2}\xi^{5}+\frac{35}{512}\chi^{4}\xi^{3}+\frac{21}{8192}\xi^{7}\right)\\ \nonumber \\
{\mathcal F}_{4}&=& f_{4}\frac{\xi^{4}}{128} + f_{5}\frac{5}{256}\chi\xi^{4} + f_{6}\left(\frac{3}{1024}\xi^{6}+\frac{15}{512}\chi^{2}\xi^{4}\right) + f_{7}\left(\frac{21}{2048}\chi\xi^{6}+\frac{35}{1024}\chi^{3}\xi^{4}\right)\\ \nonumber \\
{\mathcal F}_{5}&=& f_{5}\frac{\xi^{5}}{512} + f_{6}\frac{3}{512}\chi\xi^{5} + f_{7}\left(\frac{21}{2048}\chi^{2}\xi^{5}+\frac{7}{8192}\xi^{7}\right)\\ \nonumber \\
{\mathcal F}_{6}&=& f_{6}\frac{\xi^{6}}{2048} + f_{7}\frac{7}{4096}\chi\xi^{6}\\ \nonumber \\
{\mathcal F}_{7}&=& f_{7}\frac{\xi^{7}}{8192}
\end{eqnarray}

In a previous study~\cite{portercheb1} we showed that it was also prudent to expand the power after the log-term as a Chebyshev series.  However, in that case we had a power series to order ${\mathcal O}(v^{11})$.  In this study as the only information is given by the $v^{6}$ term, we found that there is no advantage to re-writing this term in terms of SCPs.  Again we note that $F_{C_{n}}(v) = F_{C_{n}}(v;\eta, m)$ due to the dependence of the SCPs on $v_{ini}$.  In Fig~(\ref{fig:fluxcoeffcompar}) we again plot the absolute value of the flux coefficients for both the PN and Chebyshev approximations.  We can see in this case that the PN coefficients (left) display large oscillations in their values as we go to higher PN orders.  On the other hand, we once again see that the Chebyshev coefficients show convergent properties by getting smaller as we increase the order of approximation.  We should point out that there is a slight oscillation in the values of the Chebyshev coefficients for the $\eta = 0.16$ case.  The reason for this is not fully understood.  What we have worked out is that this behaviour only exists for reduced mass ratios in the range $0.16 \leq \eta \leq 0.19$.

As in the case of the energy function, we need to have some way of comparing both flux approximations.  To this end, we use another hybrid scheme where we use the test-mass numerical flux from BH perturbation theory mixed with the $\eta$ correction terms from PN theory.  Thus the 3.5-PN fiducial exact flux is given by
\begin{eqnarray}
F_{X}(v) &=& F_{N} \left[F_{num}^{NN}(v)-\frac{35}{12}\eta v^{2} +\left(\frac{9271}{504}\eta + \frac{65}{18}\eta^{2}\right)v^{4}-\pi\frac{584}{24}\eta v^{5}\right.\nonumber\\
&+&\left.\left(-  \left(\frac{41\pi^{2}}{48}-\frac{134543}{7776}\right)\eta -\frac{94403}{3024}\eta^{2} -\frac{775}{324}\eta^{3}\right)v^{6}+\left(\frac{214745}{1728}\eta + \frac{193385}{3024}\eta^{2}\right)v^{7}\right],
\label{eqn:fx}
\end{eqnarray}
where $F_{num}^{NN}(v)$ is the Newtonian normalized numerical GW flux function~\cite{Shibata}.

In Fig~(\ref{fig:fluxcomparbhbh}) we again plot the relative error between the PN and Chebyshev GW flux approximations and the fiducial flux function for the BH-BH systems of masses $(5,5),(10,10), (20,10)$ and $(20,5)\,M_{\odot}$.  We can see from the cells in the left hand column a known trait of the PN expansion, that going to higher orders does not necessarily give a better approximation.  In any event, it is clear that in all cases the 2.5-PN flux approximation is only marginally better than the 1-PN case.  We should point out here that the 1.5 and 2-PN fluxes look much better than any other order as they approach the LSO.  This is due to a coincidental crossing of the exact flux function as the LSO is reached.  While the 3 and 3.5-PN approximations are much better earlier on, it is clear that their relative errors are greater than the 1.5 and 2-PN orders at the LSO.  However, we should point out that for the $(20,5)\,M_{\odot}$ case, the 3-PN approximation has a crossing as the LSO approaches, thus giving a very small error in the most important region.  The PN approximations can give an extremely good fit in a very narrow range as the approximated flux crosses the exact flux.  However, it is clear from the plot that where this happens for each PN order depends on the total mass of the system.

The Chebyshev fluxes, on the other hand, display a remarkable convergence.  In all cases we see oscillating error curves indicating that the Chebyshev approximations cross the exact flux many times.  While the 1-PN case has the largest amplitude of error oscillation, all other orders of approximation are incredibly consistent.  We can see that for the equal mass cases, the Chebyshev fluxes at all orders have a smaller error than PN fluxes at the 3 and 3.5-PN order.  It is clear to see from the slopes of the error curves, that the Chebyshev approximation attempts to find an equal amplitude error curve by leaving the error float in some parts in order to improve the performance elsewhere.  While it does not fully succeed, it is pretty much impossible to make out any difference between the Chebyshev approximations at PN orders of greater than one.  While the Chebyshev approximations do not have the small range accuracy of the PN approximations, overall the Chebyshev approximations display a higher stability.

\section{Fitting Factors.}\label{sec:fitfac}
We define the overlap between two waveforms $h(t)$ and $s(t)$ as the inner product of the normalized waveforms denoted by
\begin{equation}
{\mathcal O} = \frac{\left<h\left|s\right>\right.}{\sqrt{\left<h\left|h\right>\right.\left<s
\left|s\right>\right.}},
\end{equation}
where the scalar product is defined by 
\begin{equation}
\left<h\left|s\right.\right> =2\int_{0}^{\infty}\frac{df}{S_{h}(f)}\,\left[ \tilde{h}(f)\tilde{s}^{*}(f) +  \tilde{h}^{*}(f)\tilde{s}(f) \right].
\label{eq:scalarprod}
\end{equation}
Here, the * denotes complex conjugate and $\tilde{h}(f),\, \tilde{s}(f)$ are the Fourier transforms of $h(t),\, s(t)$.  The above scalar product is weighted by the inverse one-sided noise power spectral density (PSD) of the detector $S_{h}(f)$.  For initial LIGO, the design study PSD \cite{LIGO} is given by
\cite{DIS2}
\begin{equation}
S_{h}(f) = 9\times10^{-46}\left[ 0.52 + 0.16x^{-4.52} + 0.32 x^{2} \right]\ {\rm Hz}^{-1},
\end{equation}

\begin{figure}[t]
\begin{center}
\epsfig{file=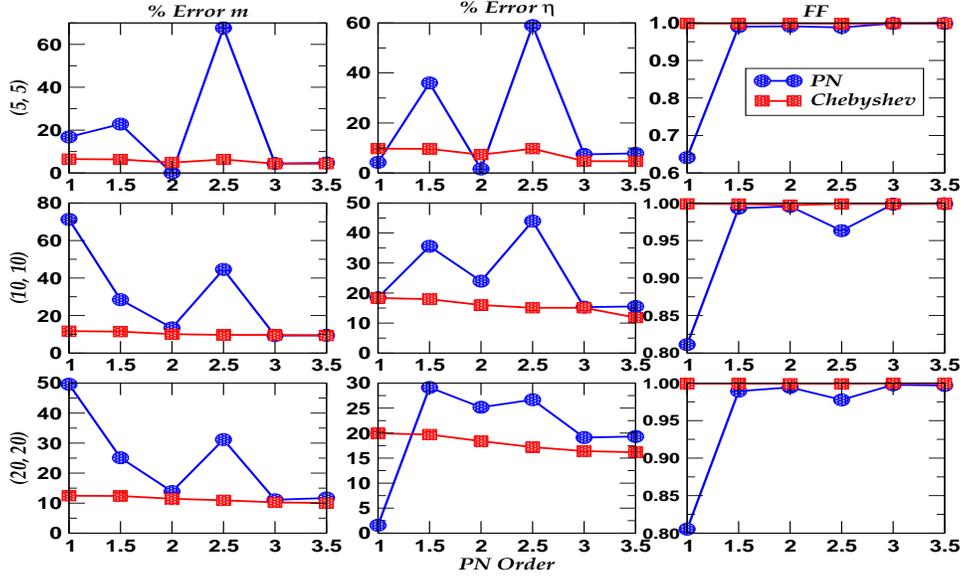, width=5in, height=3in}
\vspace{5mm}
\caption{The percentage errors in total mass $m$, reduced mass ratio $\eta$ and fitting factors for three equal mass systems of $(5,5), (10,10)$ and $(20,20)\,M_{\odot}$ for both PN and Chebyshev waveforms when compared to the fiducially exact hybrid waveform.}
\label{fig:ffem}
\end{center}
\end{figure}

\begin{figure}[h]
\begin{center}
\epsfig{file=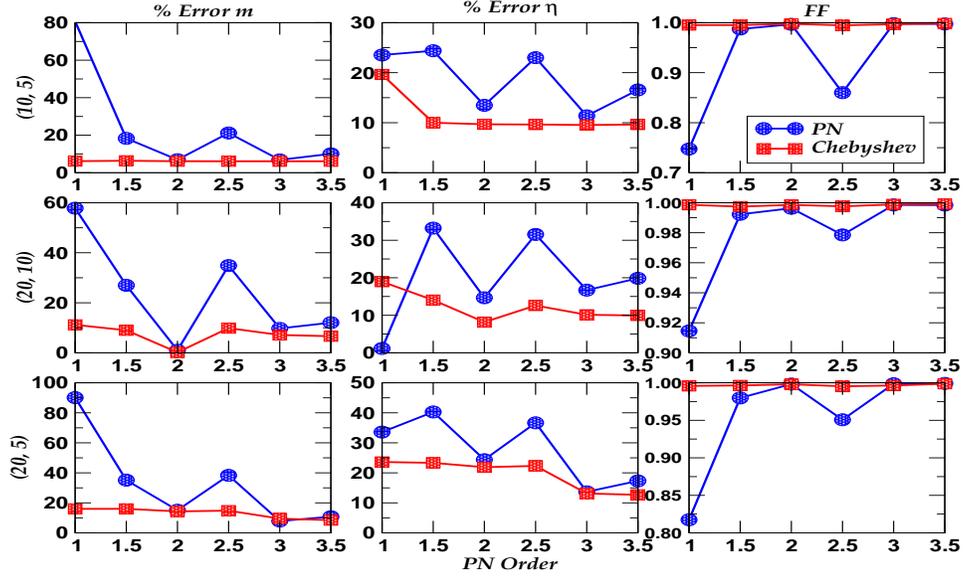, width=5in, height=3in}
\vspace{5mm}
\caption{The percentage errors in total mass $m$, reduced mass ratio $\eta$ and fitting factors for three unequal mass systems of $(10,5), (20,10)$ and $(20,5)\,M_{\odot}$ for both PN and Chebyshev waveforms when compared to the fiducially exact hybrid waveform.}
\label{fig:ffuem}
\end{center}
\end{figure}

\noindent where $x\equiv f/f_k,$ and $f_{k} = 150$\,Hz is the ``knee-frequency'' of the detector.  We take the PSD to be infinite below the lower frequency cutoff of $f_{\rm low} = 40$\,Hz.  For testing the performance of a particular template, a more appropriate quantity compared to the scalar product is the fitting factor $FF$.  Defining each template as a function of the intrinsic parameters $\lambda^{\mu}=\{m, \eta, t_{0}, \phi_{0}\}$, the fitting factor is defined as the maximum overlap obtained by varying the parameters of the template relative to the signal we are trying to detect.
\begin{equation}
FF = \max_{\lambda^{\mu}} {\mathcal O}\left(\lambda^{\mu}\right).
\end{equation}

While we expect the comparable mass waveforms to have LSO frequencies different to the test-mass case, we have verified that we can take the $f_{lso}$ from the test-mass case as the upper limit to the scalar product integral without any great change of results.  For this study we generate our signal by inserting the equations for the fiducially exact binding energy and flux functions given by Equations~(\ref{eqn:ex}) and (\ref{eqn:fx}) into the set of ODEs that describe the velocity and phase evolution of the waveform, i.e. $h_{X}(v)=h(E_{X}^{'}(v), F_{X}(v))$.

In Figs~(\ref{fig:ffem}) and (\ref{fig:ffuem}) we plot the percentage errors in the recovered total mass $m$ and reduced mass ratio $\eta$, as well as the recovered fitting factors for six different comparable mass systems.  For this study we have chosen three equal mass systems of $(5,5), (10,10)$ and $(20,20)\,M_{\odot}$, and three unequal mass systems of $(10,5), (20,10)$ and $(20,5)\,M_{\odot}$.
Firstly, focusing on the PN waveforms.  We can see that in all cases the 1-PN template performs very badly against the test waveform, never achieving very high fitting factors and always returning extremely high errors in the total mass (almost $90\%$ in the case of the $(20.5)\,M_{\odot}$ system.).  From 1.5-PN order onwards, we can always achieve fitting factors of greater than 0.95.  However the figures display the known oscillatory nature of the PN expansion.  We can see that the 2-PN is better than the 1.5-PN, but then the 2.5-PN order template is worse than the 2-PN template etc.  In all cases however, the 3 and 3.5-PN order templates return acceptable answers.  We should bring attention to the fact that for the $(5,5)\,M_{\odot}$ system, the 2-PN template outperforms the Chebyshev template in the extraction of the parameters, with a comparable fitting factor to the Chebyshev templates.  However, this behaviour does not repeat itself in any of the other cases we examined.

On the other hand, the Chebyshev templates display a remarkable convergence with all templates always achieving a fitting factor of greater than 0.99 .  In terms of parameter estimation, it is clear, in all cases, that the error at 1-PN order is only slightly greater than the Chebyshev errors at 3.5-PN order.  This is a consequence of the Chebyshev series attempting to minimize the maximum error and find a minimax solution.  While there is some oscillation in the error curves, it is quite small in comparison to the PN templates.  We can see from the figures, that compared to the PN templates, the Chebyshev templates always achieve comparable or smaller errors in the estimation of $m$ and $\eta$. 

\section{Cauchy Convergence.}\label{sec:cauchy}
If a sequence $\{x_{n}\}$ converges, the terms get closer and closer to the limit of the sequence as the order of the approximation increases.  However, the Cauchy criterion demands that rather than a limit, the terms get closer to each other.  To this extent, we define our Cauchy convergence as 
\begin{equation}
{\mathcal C} = \left< h_{n} | h_{n+1} \right>.
\end{equation}

In calculating the Cauchy convergence, the masses of both templates are kept the same as we only need to maximize over the extrinsic parameters $t_{0}$ and $\phi_{0}$.  Maximization over $t_{0}$ is achieved by simply computing the correlation of the template with the data in the frequency domain followed by the inverse Fourier transform. This yields the correlation of the signal with the data for all time-lags. To maximize over $\phi_{0}$ we use the definition of the "Best Possible Overlap" when individually maximizing over the phases of two separate templates~\cite{DIS1}.  This is given by
\begin{equation}
\left(\cos\theta_{AB}\right)_{max}=\left[\frac{A+B}{2} + \sqrt{\left(\frac{A-B}{2} \right)^{2} + C^{2}}\, \right]^{\frac{1}{2}},
\end{equation}
where
\begin{eqnarray}
A &=& \left<e_{1}^{A}\left|e_{1}^{B}\right>\right. + \left<e_{1}^{A}\left|e_{2}^{B}\right>\right.,\nonumber\\
B &=& \left<e_{2}^{A}\left|e_{1}^{B}\right>\right. + \left<e_{2}^{A}\left|e_{2}^{B}\right>\right., \\
C &=&  \left<e_{1}^{A}\left|e_{1}^{B}\right>\right.\left<e_{2}^{A}\left|e_{1}^{B}\right>\right. + \left<e_{1}^{A}\left|e_{2}^{B}\right>\right.\left<e_{2}^{A}\left|e_{2}^{B}\right>\right. ,\nonumber
\end{eqnarray}
and
\begin{equation}
e_{1}^{A} = \frac{\tilde{h}_{1}^{A}}{|\tilde{h}_{1}^{A}|},\ \ \ \ 
e_{2}^{A} = \frac{\tilde{h}_{2}^{A} - \left<h_{2}^{A}\left|e_{1}^{A}\right>\right.e_{1}^{A}}{|\tilde{h}_{2}^{A} - \left<h_{2}^{A}\left|e_{1}^{A}\right>\right.e_{1}^{A} |},
\end{equation}
where $\tilde{h}_{1}^{A} = \tilde{h}\left(\phi_{0} = 0\right)$ and $\tilde{h}_{2}^{A} = \tilde{h}\left(\phi_{0} = \pi / 2\right)$.

The Cauchy convergence has been calculated many times before for PN templates (See for example~\cite{DIS1}), so we will not repeat the exercise here.  Suffice to say that the PN templates display the usual oscillatory behaviour which shows that going from one PN order to the next does not necessarily result in a better template.  In Table~(\ref{tab:ccem}) we list the Cauchy convergence for the Chebyshev templates for the six test cases we analysed in the previous sections.  The parameter $n$ denotes the level of approximation, e.g. $n=2$ corresponds to 1-PN or $v^{2}$.  We can see from the tables that the Chebyshev templates are incredibly convergent with all templates achieving overlaps of greater than 0.9 with the successive template.  In fact it is only in the $(5,5)\,M_{\odot}$ case that the template fails to achieve overlaps of $\geq 0.97$ with each other.

\begin{table}[t]
\begin{tabular}{|c|c|c|c|c|c|c|}\hline \hline\hline
 $n$ & $(5, 5)M_{\odot}$ & $(10, 10)M_{\odot}$ & $(20, 20)M_{\odot}$ & $(10, 5)M_{\odot}$ & $(20, 10)M_{\odot}$ & $(20, 5)M_{\odot}$\\ \hline\hline
2 & 0.9497 & 0.9841 & 0.99991  & 0.9703 & 0.9998 & 0.99987   \\
3 & 0.99969 & 0.9999 & 1.0  & 0.99953 & 0.999996 & 0.99978   \\
4 & 0.9998 & 1.0 & 1.0 & 0.99949 & 1.0 & 0.99998    \\
5 & 0.9999 & 1.0 & 1.0  & 0.99999 & 1.0 & 1.0   \\
6 & 1.0 & 1.0 &  1.0 & 1.0 & 1.0 & 1.0
  \\ \hline\hline
 
\end{tabular}
\caption{The Cauchy convergence $\left< h_{n} | h_{n+1} \right>$ of the Chebyshev templates for three equal mass, and three unequal mass systems assuming a LIGO noise curve and a lower frequency cutoff of 40 Hz.}
\label{tab:ccem}
\end{table}

\section{Position of the LSO.}\label{sec:lso}
In the previous sections of this work we took the position of the LSO to be at $R=6\,m$ thus giving the dimensionless velocity of $v_{lso}=1/\sqrt{6}$.  Here we investigate how close both the PN and Chebyshev approximations come to correctly predicting the position of the LSO, defined as $E'(v) = 0$.  For both the `exact' and PN expansions this quantity is $E'(v;\eta) = 0$, while for the Chebyshev approximation it is $E'(v;\eta,m) = 0$ giving different values for different total masses.

\begin{figure}[t]
\vspace{0.25 in}
\begin{center}
\centerline{\epsfxsize=12cm \epsfysize=8cm \epsfbox{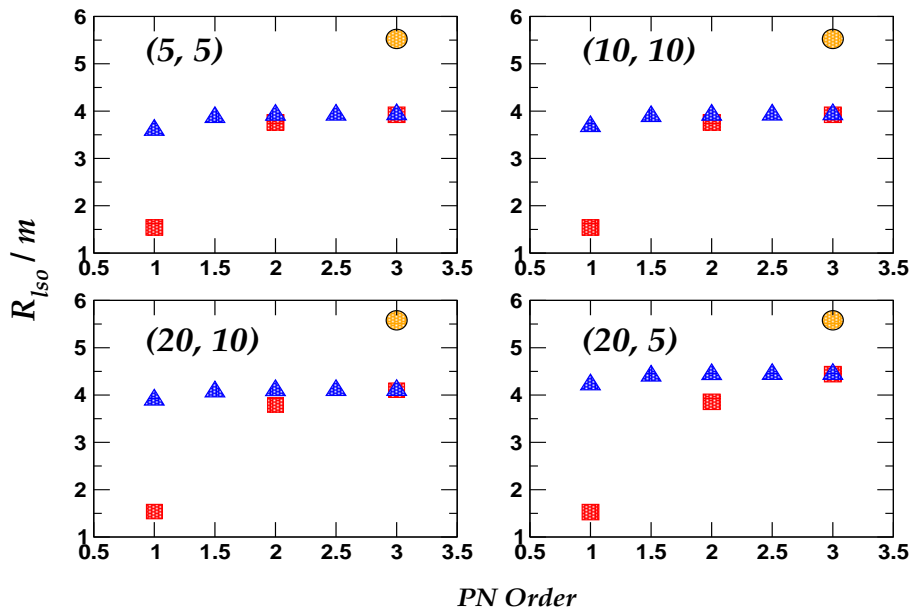}}
\vspace{5mm}
\caption{Comparison of the position of the LSO for the PN and Chebyshev energy functions against the minimum of the fiducial exact energy function for the four BH-BH systems with masses $(5,5),(10,10), (20,10)$ and $(20,5)\,M_{\odot}$.  The circles denote the LSO for the exact fiducial energy, the squares represent the PN approximation and the triangles represent the Chebyshev approximation.  We can see that the Chebyshev approximation is more convergent and gives a better prediction of the location of the LSO at virtually all PN orders.}
\label{fig:lsocompar}
\end{center}
\end{figure}

In Fig~(\ref{fig:lsocompar}) we plot the positions of the predicted LSOs against the fiducial value for four of the BH-BH systems that we used previously.  In each cell, the circle denotes the value of the exact LSO position in units of $R/m$.  We have placed the exact position of the LSO at 3-PN to signify the order of $\eta$ corrections involved.  The squares denote the PN approximation values and the triangles denote the Chebyshev approximations.  

In the top two cells we use the same values for both the fiducial and PN predicted LSOs.  These cells correspond to the $\eta = 0.25$ case which does not change for the PN approximation, but does for the Chebyshev approximation.  We can see that in all four cases, the PN approximation at 1-PN gives a very bad prediction of the position of the LSO.  This value is improved somewhat at 2-PN order, with the best value coming in all cases from the 3-PN order PN approximation.

Just like other re-summation methods, there is no new information gained from restructuring the series approximation.  Thus, there are no obvious tricks we can use to ensure that the location of the LSO is improved at the highest PN order.  Therefore, it is no surprise that the PN and Chebyshev templates predict the same location at 3-PN order. However, we can see from the image that the prediction for the LSOs position is greatly improved at lower PN orders with the Chebyshev templates.  The flat profile of the error distribution is consistent with the effort of the Chebyshev approximation to minimize the maximum error.

\begin{figure}[t]
\begin{center}
\epsfig{file=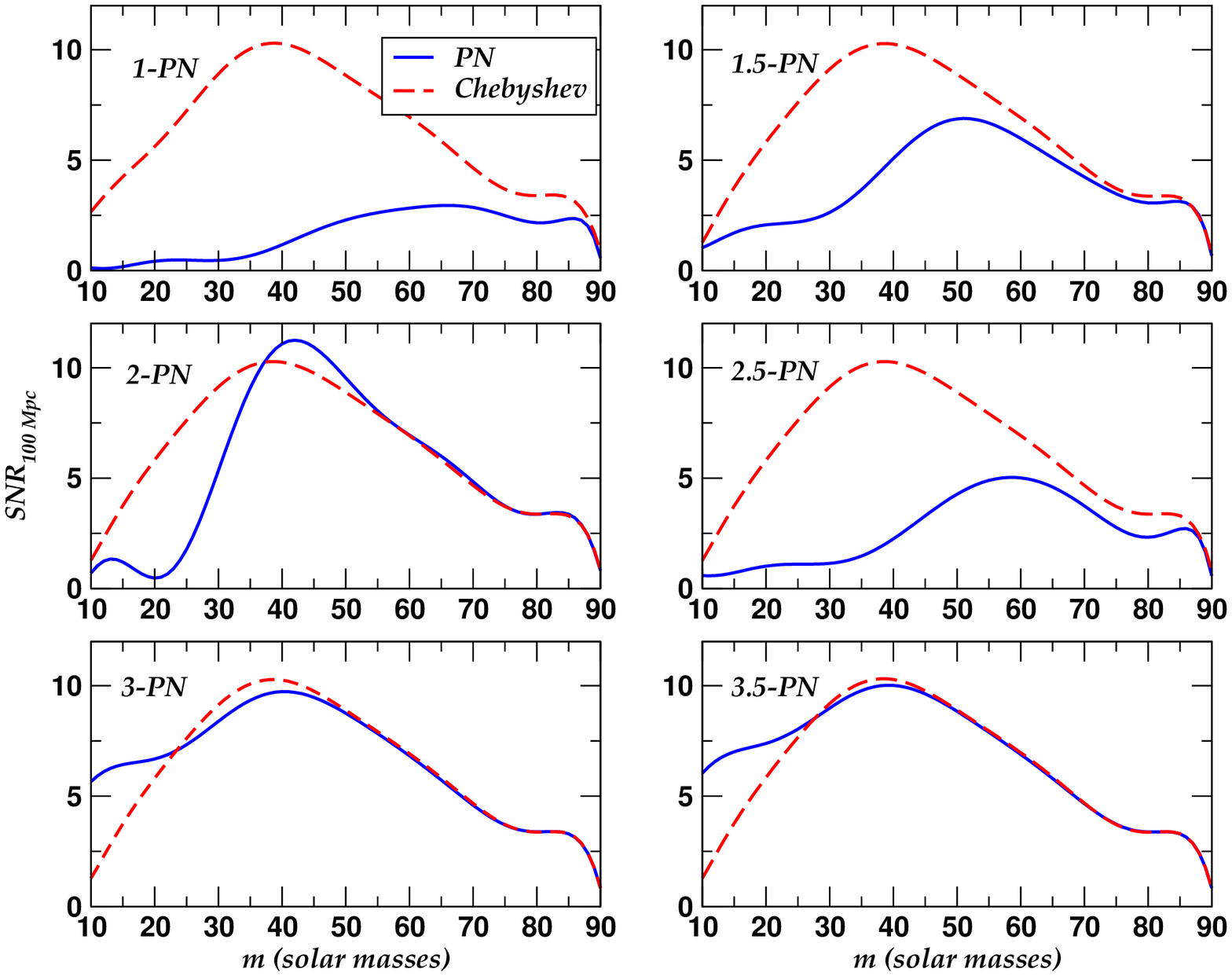, width=5in, height=3in}
\vspace{5mm}
\caption{Comparison of the recovered SNRs for both PN and Chebyshev templates using the fiducially exact signal for equal mass systems.  The sources are set at a common distance of 100 Mpc.}
\label{fig:snrcompar}
\end{center}
\end{figure}

\section{Signal-To-Noise Ratios.}\label{sec:snr}
The final aspect we will investigate is a comparison of recovered signal-to-noise ratios (SNRs) using the fiducially exact template as our signal.  Using the definition of the scalar product, we can write the signal-to-noise ratio (SNR) obtained by searching for a signal $h_{X}(t)$ with a template $h_{n}(t)$ as 
\begin{equation}
\rho = \frac{\left<h_{n}\left|h_{X}\right>\right.}{\sqrt{\left<h_{n}\left|h_{n}\right>\right.}}
\end{equation}
For this investigation we place the exact waveform and template at a distance of 100 Mpc. We also assume that the parameters of the template match the parameters of the signal exactly.  In Fig~(\ref{fig:snrcompar}) we plot the recovered SNRs at various PN orders for equal mass systems.  The results scale as $\sqrt{4\eta}$ for unequal mass binaries.  We can see from the plots that at 1, 1.5 and 2.5-PN orders the Chebyshev templates recover a much greater SNR than the PN templates at virtually all masses.  At the 2-PN order the Chebyshev templates outperform the PN templates at low masses and are then comparable at the higher masses.  While for the 3 and 3.5-PN orders, the PN templates outperform the Chebyshev templates at very low masses, but are then comparable at higher masses.

\section{Conclusion.}
In this work we have presented a new family of templates to detect GWs from the inspiral of comparable mass BH-BH binaries.  The templates are based on shifted Chebyshev polynomials which are bound between the initial and final velocities of the system in question.  In general, the Chebyshev polynomials display the fastest convergence of all the orthogonal polynomials.  Using the shifted polynomials, we have defined a new re-summed binding energy and GW flux function which we demonstrated to be graphically more convergent than the PN functions when compared with a fiducially exact energy and flux.  Using a fiducially exact template constructed from the exact energy and flux, we were able to show that the Chebyshev templates achieve very high fitting factors and excellent parameter extraction as compared to the PN templates.  We were also able to show that in an effort to get close to a minimax solution, the 1-PN Chebyshev template is almost as good as the 3.5-PN PN template.  Further displaying the strength of these new templates, we were also able to show that they provided more accurate measurements for the position of the LSO, have an incredibly fast Cauchy convergence, and in most cases achieve a higher recovered SNR.  

We should point out that the Chebyshev templates outperformed the PN templates when compared to a fiducially exact waveform constructed from a hybrid combination of the test-mass binding energy and a numerical flux function from BH perturbation theory.  These functions were then combined with the mass dependent parts of the PN expansion for the energy and flux functions for comparable-mass bodies.  While is seems sensible, as we recover the test-mass result in the limit $\eta\rightarrow 0$, it is still only an approximation of a true possible waveform.  However, we are confident in the abilities of this new template family.  We plan to follow up this investigation with a comparison against comparable-mass waveforms from numerical relativity.  This should provide us with a more concrete critique of the ability of these new Chebyshev templates to detect GWs and parametrize their sources.

For now, we believe that this new family of comparable-mass templates will be a valuable addition to the search for BH-BH binaries with both ground and space-based detectors.

\end{document}